\documentstyle[twocolumn]{article}

\begin{document}
\onecolumn

\title{$4d$ Simplicial Quantum Gravity
       with a Non-Trivial Measure}

\author{
Bernd BR{\"U}GMANN and Enzo MARINARI$^{(*)}$\\[1.5em]
Dept. of Physics and NPAC,\\
Syracuse University,\\
Syracuse, NY 13244, USA\\
{\footnotesize
  bruegmann@suhep.phy.syr.edu
  marinari@roma2.infn.it}\\[1.0em]
{\small $(*)$:  and Dipartimento di Fisica and Infn,} \\
{\small Universit\`a di Roma {\it Tor Vergata}}\\
{\small Viale della Ricerca Scientifica, 00173 Roma, Italy}}
\date{October, 1992}
\maketitle

\begin{abstract}
We study $4d$ simplicial quantum gravity in the dynamical
triangulation approach with a non-trivial class of measures. We find
that the measure contribution plays an important role, influencing the
phase diagram and the nature of the (possible) critical theory. We
discuss how the lattice theory could be used to fix the quantum
measure in a non-ambiguous way.
\end{abstract}

\vfill

\begin{flushright}
  {\bf PACS 04.60.+n}\\
  {\bf SCCS 361}\\
  {\bf SU-GP-92/9-2}\\
  {\bf hep-lat/9210002 }\\
\end{flushright}

\twocolumn

Dynamically triangulated random sur\-faces (DTRS) \cite{DYTRAS} play an
important role in the efforts to develop a coeherent description of quantum
gravity. The (euclidian) space-time is approximated by a $d$-dimensional
simplicial triangulation, where the link length is constant, equal to $1$,
but the connectivity matrix is a dynamical variable.

The most important advances have been obtained in two-dimensional
quantum gravity, where DTRS are simplicial triangulations of a $2d$
manifolds. The analytic success of matrix models, which can be for
example exactly solved in the case of pure $2d$ gravity \cite{EXACT},
has strongly encouraged this approach.  The results obtained in the
triangulated approach and in the continuum lead to consistent
predictions for correlation functions and critical exponents.

Dynamical triangulations are also potentially relevant in four
dimensions. One can hope that a sensible, non-perturbative definition
of the quantum gravity theory can be obtained in some {\em scaling}
limit of the theory of $4d$ hyper-tetrahedra. This approach has much
in common with Regge calculus, where the connectivity is fixed but the
functional integration runs over the link lengths. The underlying
principle is clearly very similar, and one could say that DTRS have
the status of an improved Regge calculus.  We face the usual problem
inherent in discretizing a theory, i.e. the discretization scheme can break
some of the continuous symmetries, which will have to be recovered in
the continuum limit (if there is one). Indeed, Wilson lattice gauge
theories have taught us an important lesson. The fact that gauge
invariance is exactly conserved in the lattice theory, for all values
of the lattice spacing $a$, is in that case crucial: it would have
been very difficult to establish firm numerical results if one would
have had to care about the presence of non gauge-invariant correction,
which would disappear only in the $a \to 0$ limit. In the case of
quantum gravity, diffeomorphism invariance plays such a crucial role,
and DTRS are diffeomorphism invariant by construction, at least on the
space of piecewise flat manifolds. Hence part of the difficulties Regge
calculus has in forgetting about the lattice structure are eliminated
a priori in the DTRS lattice approach.

There are two more important points to stress. The first one is that
in the DTRS approach in $3d$ and $4d$, as opposed to $2d$, we can try
to make sense out of the pure Einstein action, without, for example,
curvature squared terms. Even though the partition function formally
diverges, at fixed volume the local curvature is bounded both from
below and from above. Therefore we can study the theory at fixed (or
better quasi-fixed, see later) volume, and look for the existence of a
stable fixed point in the large volume limit. A second order phase
transition with diverging correlation lengths, in the statistical
mechanics language, would allow us to define a continuum limit which
is universal and is not influenced by the details of the underlying
discrete lattice structure. Precisely this scenario constitutes one of
the best hopes we have to find a consistent quantum theory of gravity.
If euclidean quantum gravity based on the Einstein action does have
non-perturbative meaning, then we can exhibit it in this way.

The second problem is to determine the measure one should use to
define the quantum theory. This problem, far from being solved in the
continuum, is completely open in the lattice approach (for a good
review see for example \cite{MENOTT}). In quantum Regge calculus the
influence of the measure has been examined in \cite{BGM} but there is
no direct relation to DTRS. The problem of the measure is the main
point we address in this note, and we want to suggest that the DTRS
approach may be powerful enough to solve it.

Recently two groups have pionereed Monte Carlo simulations of DTRS
\cite{AMBJUR,AGIMIA,AGIMIB}.
Numerical simulations (even if on quite a small scale) turned out to
be feasible, and lead to very non-trivial results. One clearly
observes a phase transition structure. Although quantitative
statements are not easy to make, given the limited statistics and the
small lattice size, it is clear that the situation is different from
the $3d$ case, where the phase transition is manifestly of first
order, and there is no continuum theory. In $4d$ \cite{AMBJUR} there
is an open possibility that the transition is second order (although
that cannot be claimed without a much more detailed finite size
study). One does not observe hysteresis cycles, and the crossover is
less sharp then in $3d$. More involved statements about critical
exponents have to be taken at this point, we believe, {\em cum grano
salis}, but there is evidence for the possible presence of a critical
point with a non-trivial continuum theory in the phase diagram for the
euclidean Einstein action. This observation certainly warrants further
careful investigations.

These first simulations have been run with uniform measure, where all
triangulations have the same weight in the sum which defines the path
integral of the quantum theory. There are no particularly good
reasons for this choice to be the correct one, and in the following we
investigate the changes introduced by defining simplicial quantum
gravity with a non-trivial measure.

One of us has described in \cite{BERND} the structure of the Monte
Carlo simulations, and how the efficiency of the numerical procedure
can be optimized. The programming of dynamical triangulations is
difficult since a dynamical data structure is required, but it is very
relevant in many practical applications.  For the same reason the
implementation of DTRS on parallel computers is a hard problem.  Since
a set of ergodic moves preserving the volume for canonical simulations
is not known, one has to consider a Markov chain which sweeps out the
space of different volume simplicial manifolds.  We have used the
quasi-canonical method introduced in \cite{BAUBER}, which allows us to
control the systematic distortion arising from the {\em ad hoc}
potential that keeps the system close to a given number of
$4$-simplices \cite{BERND}. All the details about the system we study
and about the numerical procedures are given in \cite{BERND}.

In the continuum the euclidean Einstein-Hilbert action for a metric
$g_{\mu\nu}$ has the form
\begin{equation}
  \protect\label{E_EH}
  S_E[g] = \int d^4 x \sqrt{g}(\lambda-\frac{R(g)}{G})\ ,
\end{equation}
where $R(g)$ is the Ricci scalar and $\lambda$ and $G$ are the
cosmological and gravitational constants, respectively. We consider a
fixed $S^4$ topology.  On a
triangulation $T$ we discretize according to
\begin{equation}
  V = \int d^4 x \sqrt{g} \to N_4[T]\ ,
\end{equation}
\begin{equation}
  R = \int d^4 x \sqrt{g}R(g) \to \frac{2 \pi}{\alpha} N_2[T] - 10 N_4[T]\ ,
\end{equation}
where $\alpha\simeq 1.318$, and $N_i[T]$ is the number of $i$-simplices of
the triangulation $T$. The discrete action is then
\begin{equation}
  \protect\label{E_EHL}
  S_E[T] = k_4 N_4[T] - k_2 N_2[T]\ ,
\end{equation}
where $k_4 = \lambda + 10/G$ and $k_2 =  2\pi / \alpha G$.

In the discrete quantum theory there exists a critical line
$k_4=k_4^c(k_2)$ such that if $k_4$ is different from $k_4^c$ for a
given value of $k_2$ then the random walk tends to either zero or
infinite volume. All measurements are made for $k_4=k_4^c(k_2)$.

We have selected not one but a family of measures in order to
investigate the influence of the measure in a rather general setting.
Our choice is guided by diffeomorphism invariance of the measure
\cite{MEASURE} but ignores more sophisticated arguments like BRST
invariance. We have studied, as a function of $n$, a measure
contribution of the form
\begin{equation}
  \prod_x g^{n/2}\ ,
\end{equation}
i.e. in the triangulated theory $S_E[T]$ is replaced by $S[T] = S_E[T]
+ S_M[T]$, where
\begin{equation}
  S_M = - n \sum_a \log \frac{o(a)}{5}\ .
\end{equation}
The sum runs over all $0$-simplices (sites) of the manifold, and
$o(a)$ is the number of $4$-simplices which include the site $a$. We
considered $n$ in the interval from $-5$ to $5$. The case $n=0$
repeats simulations with the trivial, uniform measure, which can be
compared with previous results.

Let us summarize our results. We confirm the fact that the phase
transition can be of second order, and that it is plausible that we
will be able to define a sensible theory. We find that the measure
factor plays an important role, and that the critical behavior does
depend on $n$. This is very different from $2d$ quantum gravity (see
for example \cite{BCHHM}), where modifications of the measure factor
of the same kind we use here do not have any non-trivial effect on the
critical behavior. Varying $n$ does not only change non-universal
quantities, like for example the value of the critical coupling, but
changes the actual (pseudo-)critical behavior.

In figure 1 we plot the average curvature $R/V$ for $V\equiv N_4=4000$
as a function of the coupling $k_2$ for different values of $n$,
$n=-5$ for the lowest curve, then $n=-1$, $0$, $1$ and $n=5$ for the
upper curve. In figure 2 we plot the average distance (in the internal
space) of two 4-simplices. We count the minimum number of steps from
4-simplex to 4-simplex across 3-simplex faces that connect a pair of
4-simplices and average over all 4-simplices and random manifolds.

Both figures show that the measure operator has a pronounced effect.
Increasing the coupling of the measure term leads to a continuous,
monotonous deformation of the curves. Notice that the curves are not
just shifted. In the case of $R/V$, the singularity seems stronger for
$n \simeq 0$, where the jump in $R/V$ is quite sharp. The distance $d$
has a sharper jump for $n=1$, where it seems to jump from one constant
value to another constant. Smaller values of $n$ show a slower
increase in $d$.

For large absolute values of $n$, especially for $n=-5$, the plots
show a weaker singularity. The profile of $R/V$ hints less at a sharp
jump than the former cases, and the distance increases very smoothly
from a critical value of $k_2$, $k_2^c(n)$ on. When $n$ increases to
the value of $5$ the system seems to loose criticality on an absolute
scale.  Its behavior through the crossover is quite smooth.

A critical value of $k_2^c$ can be defined, for example, as the point where
the distance value starts to change. But for the $n=5$ case the transition
point is not very clear. Let us note that such a value of $k_2^c$ changes its
sign as a function of $n$.

Figure 3 and 4 show results for $N_4=16000$ to indicate what kind of
finite size effects are present. As evident from figure 3, larger
volumes amplify the effect of large absolute $n$ for $k_2<k_2^c$.
Figure 4 for the average distance $d$ displays the same qualitative
behaviour as figure 2 for $k_2<k_2^c$, but for large $k_2$ the average
distance does not remain constant, which might already have been guessed
in figure 2. The explanation is that the measure term $S_M$ for
positive $n$ introduces a bias towards smaller $o(a)$ which increases
$d$, but for large enough $k_2$ and $k_4$ the contribution of $S_E$
dominates and $d$ approaches its $n=0$ value. This may be the
case for $k_2>k_2^c$, while for $k_2>k_2^c$ the critical value
$k_4^c(k_2)$ is such that $S_E\approx 0$, and at least for the range of
negative $k_2$ considered here $S_M$ is relevant and $d$ remains
large for $n=5$. The same holds for $R/V$ when $n=5$.

Our conclusion is that the measure term has a strong effect, which
seems difficult to reabsorb in a simple renormalization of the
critical coupling. Always keeping in mind that a precise finite size
study is required before making quantitative statements \cite{INPREP},
we believe there are two basic possibilities. The first possibility is
that there is only one universality class, and that all the theories
we have studied do asymptotically show the same critical behavior. In
this case the rate of approach to the continuum limit is strongly
influenced by $n$. We will select the theory with faster convergence
to the continuum.

The second possibility (which is the most interesting one) is that the
measure factor changes the universality class. Our results, albeit
preliminary, seem to hint in this direction. In this case we could
have a critical value of $n_c$, and transitions belonging to different
universality classes. This is a very appealing scenario, and here the
lattice discrete theory could make its own original contribution. It
could be possible to pick out the correct measure, on the lattice, by
requiring a particular expectation value and scaling behavior of some
physical observable. Such a prescription would be a powerful tool,
turning the discrete version of the theory from a source of
indetermination into a completely determined scheme.

\section*{Acknowledgments}

Lee Smolin has emphasized, on many occasions, the role of
the measure in a theory of quantum gravity, and we acknowledge many of
his interesting comments. B.B. was supported in part by NSF grant 90
16733. The support of Geoffrey Fox and of NPAC have been crucial to
the success of this project.

\vfill
\end{document}